\documentclass[twocolumn, preprintnumbers,amsmath,amssymb,balancelastpage]{revtex4}
\usepackage{hyperref}
\usepackage{amsmath}
\usepackage{amssymb}
\usepackage{graphicx}
\usepackage{slashed}
\usepackage{dcolumn}
\usepackage{bm}
\usepackage[inline]{enumitem}
\usepackage{xcolor}
\usepackage{dsfont}
\usepackage{verbatim}
\usepackage{mathtools, xcolor,ytableau, amsfonts,tikz}
\usepackage{tabularx}
\usepackage{putexPRL}

\begin{document}

\preprint{}

\title{Symmetry Enhancement, SPT Absorption, and Duality in QED$_3$}

\author{Shai M. Chester}
\affiliation{Blackett Laboratory, Imperial College, Prince Consort Road, London, SW7 2AZ, U.K.}

\author{Zohar Komargodski}
\affiliation{Simons Center for Geometry and Physics, SUNY, Stony Brook, NY 11794, USA}

\date{\today}

\begin{abstract}

Quantum Electrodynamics in 2+1 dimensions (QED$_3$) with two Dirac fermions displays time reversal symmetry, nontrivial SPT phases and anomalies. The fate of this theory in its strongly coupled regime has been debated extensively. Surprisingly, we find that gluing together the phase diagrams of two standard Wilson-Fisher $O(4)$ theories suffices to reproduce all the SPT phases, anomalies, and semi-classical limits. A central mechanism behind it is ``SPT absorption''. The patching of the $O(4)$ transitions makes very concrete predictions for the behavior of the theory in its strongly coupled limits; for instance, the $\theta=\pi$ sigma model with $S^3$ topology appears due to monopole condensation.

\end{abstract}

\maketitle

\section{Introduction}

A $U(1)$ gauge field interacting with Dirac fermions in 2+1 space-time dimensions leads to a class of interesting, nontrivial, and experimentally relevant continuum Quantum Field Theories (QFTs). Among them, there are the special theories that preserve time-reversal invariance for massless fermions. For brevity, we will refer to these as the QED$_3$ theories. The Lagrangian at the massless point is simply given by
\begin{equation}\label{kinetic}L_{kinetic}=-{1\over 4e^2}F^2 + i \bar \Psi_i \slashed{D} \Psi^i~, \end{equation} 
where $i=1,...,N_f$ is the number of Dirac fermions species. For this theory to exist we must take $N_f$ to be an even number (this is in order to avoid the breaking of time reversal symmetry by fermion loop diagrams~\cite{Redlich:1983kn})\footnote{ Throughout, we do not deform the action by monopole operators, and therefore, our theory will always have a $U(1)$ symmetry associated to the conservation of the total magnetic charge (this symmetry may or may not be broken spontaneously, though).}.

One basic question is to understand the low-energy physics and determine the phase diagram as a function of the fermion mass matrix $ M_{i}^{j}\bar \Psi^i \Psi_j$. For sufficiently large $N_f$, calculations are possible in a $1/N_f$ expansion (as first demonstrated in~\cite{Appelquist:1988sr}). There is a single second order transition at $M=0$ and for nonzero $M$ there are quantum Hall phases (as well as a phase with a $U(1)$ Nambu-Goldstone boson (NGB) due to monopole condensation~\cite{Polyakov:1976fu} when the number of positive and negative eigenvalues of $M_{i}^{j}$ coincide). 

For small values of $N_f$, and in particular for the minimal nontrivial case $N_f=2$, there are several speculations. One scenario is that the global $SU(2)\times U(1)$ symmetry is spontaneously broken \footnote{Using RG monotones, there exist rigorous bounds on symmetry breaking phases, see e.g.~\cite{Metlitski:2011pr,Grover:2012sp,Sharon:2018apk}.}.
A symmetry breaking pattern that was discussed extensively in the literature is 
 $SU(2)\times U(1)\to U(1)\times U(1)$ which leads to 2 NGBs, and is due to the condensation of the mass operator $\bar \Psi^i \Psi_j$ \footnote{A related scenario for non-compact QED$_3$ with symmetry breaking pattern $SU(2)\to U(1)$ has received support from lattice simulations \cite{Hands:2020itv,Hands:2002dv,Strouthos:2007stc}.}. Another more recent scenario postulates the existence of a 2nd order transition with symmetry enhancements~\cite{Karch:2016sxi,Seiberg:2016gmd,Wang:2017txt,Akhond:2019ued}, see also the review~\cite{Senthil:2018cru}.
These symmetry enhancement scenarios at a conformal fixed point are implausible in light of the bootstrap results~\cite{Li:2021emd} (and see references therein). Related work can be found in~\cite{Karthik:2015sgq,Karthik:2016ppr,Qin:2017cqw,Serna:2018tct}.

The impetus for this paper is a remarkable similarity between the scaling exponents of scalar operators in the rank $q$ traceless symmetric irrep of $O(4)$ in the Wilson-Fisher (WF) fixed point, and charge $q$ monopole operators in $N_f=2$ QED$_3$ \footnote{We normalize the monopoles to have charge $\int_{S^2}F=2\pi q$, so $q\in\mathbb{Z}$, which is different from the normalization used in \cite{Pufu:2013vpa,Dupuis:2021flq,Dyer:2013fja}.}, which transform in the corresponding irrep upon the decomposition $SO(4)\to SU(2)\times U(1)$. The computation on the QED$_3$ side is via a $1/N_f$ expansion to subleading order \cite{Pufu:2013vpa,Dupuis:2021flq,Dyer:2013fja}, which was previously shown to be accurate for small values like $N_f=4$ by comparison to conformal bootstrap results \cite{Chester:2016wrc,Albayrak:2021xtd} \footnote{In the related case of a $U(1)$ gauge field coupled to $N_f$ complex scalars, the large $N_f$ calculation of monopole operator scaling dimensions was shown to be accurate for all $N_f$ \cite{Dyer:2015zha} by comparison to lattice data, and even for $N_f=1$ by comparison to particle/vortex duality \cite{Chester:2022wur}. Similarly, the large $N_f$ and Chern-Simons level $k$ results \cite{Chester:2017vdh,Chester:2021drl} were matched to the 3d bosonization duality for $N_f=k=1$ \cite{Chester:2022wur}.}. The $O(4)$ WF scaling dimensions were computed by a lattice simulation \cite{Banerjee:2019jpw}. In the Appendix, we see that the corresponding operators match for all values of $q$, and the match even gets better as $q$ grows \footnote{{In the large charge limit, we expect the scaling dimenions are controlled by a few Wilson coefficients and behave as $\sim q^{3/2}$ to leading order~\cite{Hellerman:2015nra,Dupuis:2021flq}.}}. Furthermore, if one employs the web of dualities~\cite{Aharony:2015mjs,Seiberg:2016gmd} away from its regime of validity then one finds a duality between $N_f=2$ QED$_3$ and $O(4)$ WF \footnote{In particular, this follows from plugging in $N_f=2,N=1,k=1$ into the first duality in equation $1.4$ of \cite{Seiberg:2016gmd}.}. Such a duality was actually suggested earlier in~\cite{alicea2005criticality}.

However, {despite these tantalizing numerics}, it cannot be that the massless QED$_3$ model flows to the $O(4)$ WF theory for several reasons, most importantly, because there is a 't Hooft anomaly involving time reversal symmetry in QED$_3$. Another reason is that the relevant singlet in QED$_3$ is $\bar \Psi_i \Psi^i$, which is time-reversal odd, while the relevant singlet in the WF theory is time-reversal even. Finally, the WF theory cannot match the SPT phases that we will explain later.

\begin{figure}[!h]
  \includegraphics[width=0.51\textwidth ]{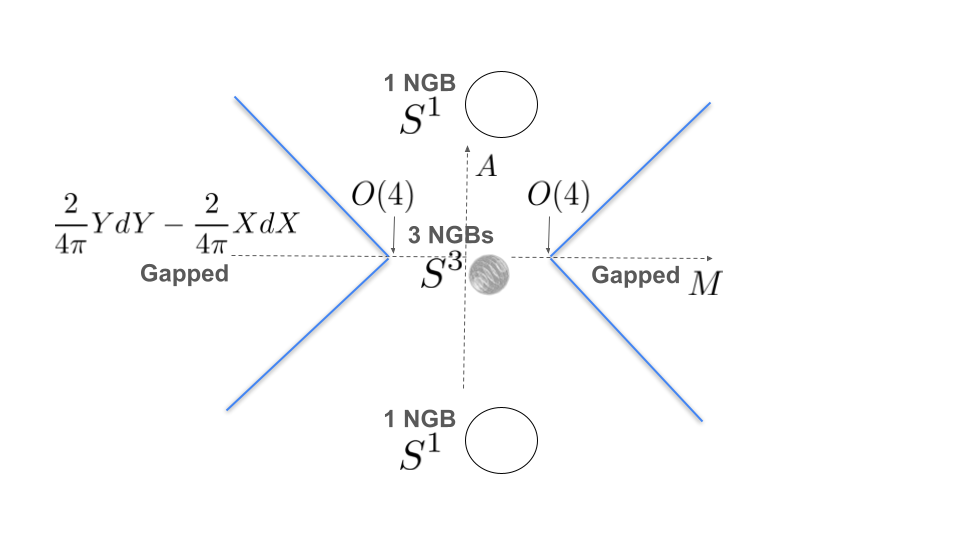}
  \caption{The proposed phase diagram of QED$_3$. The horizontal axis corresponds to the $SU(2)\times U(1)$ preserving mass term while the vertical axis corresponds to the mass term in the adjoint representation of $SU(2)\times U(1)$. {We denote these two types of mass terms by $M$ and $A$, respectively.} On the horizontal axis, in the symmetry broken phase, the theta angle of the sigma model varies continuously from 0 to $2\pi$, with  $\theta=\pi$ at $M=0$. The $\pm 45$ degrees lines correspond to $O(2)$ Wilson-Fisher transitions and these lines meet at two $O(4)$ Wilson-Fisher transitions.  }\label{FinalD}
\end{figure}

We will show that gluing {\it two} $O(4)$ WF theories can reproduce the SPT phases and the anomaly of QED$_3$! 
A crucial ingredient is that some SPT phases can be absorbed in the symmetry broken phase (SPT absorption). Another important ingredient is that a sigma model with $S^3$ target space can have a $\theta$ angle in 2+1 dimensions. Our picture clarifies in what sense QED$_3$ and $O(4)$ are dual and it suggests an explanation for why there is such a remarkable agreement between scaling exponents. The large $N_f$ expansion of the monopole scaling dimensions is performed in the regime where the massless theory flows to a CFT, while we are extrapolating to $N_f=2$ where we claim the massless theory experiences spontaneous symmetry breaking, but the massive theory flows to the $O(4)$ CFT. It could be that this extrapolation happens to land on the $O(4)$ transition. Or it could be that the massless theory is very close in RG space to QED$_3$ with massive fermions, which is dual to $O(4)$ WF.

The scenario we describe can be viewed as an extension of the ``quantum phase'' paradigm in~\cite{Komargodski:2017keh, Gomis:2017ixy, Choi:2018tuh,Armoni:2019lgb,Argurio:2020her}.
First of all, we propose that the massless theory breaks the symmetry as $SU(2)\times U(1)\to U(1)$ due to condensation of the $q=1$ (elementary) monopole,
\begin{equation}\label{CondM}\langle \mathcal{M}\rangle\neq 0\end{equation} 
leading to 3 NGBs (unlike the previously discussed symmetry breaking pattern, due to the condensation of a fermion bilinear). These 3 NGBs live on a target space of $S^3$ topology with a squashed metric and with $\theta=\pi$ at the massless point of QED$_3$, leading to the infrared effective action~\eqref{kinetic}
\begin{align}\label{infraredmassessl}& S^{ir}_{ M_i^j=0} =\frac{f}{2} \int d^2xdt \ \delta^{ab}\text{Tr}(\partial_a g \partial_b g^{-1})+{\rm squashing}\nonumber\\&+{1\over 24\pi} \int d^2xdt\ \epsilon^{abc}\text{Tr}(g^{-1}\partial_a gg^{-1}\partial_b gg^{-1}\partial_c g)~,\end{align}
where $g$ is an $SU(2)$ matrix and the squashing term breaks the symmetry of the nonlinear sigma model to $SU(2)\times U(1)$ by squashing the target space metric.

As we turn on masses $M_i^j$, the geometry of the $S^3$ slightly changes and the $\theta$ angles change as well. Indeed, the mass term $M_i^j\sim \delta^i_j$ preserves $SU(2)\times U(1)$ and is time-reversal odd. Hence the map between the operators at low energies near the massless point is \footnote{The map for other operators that are charged under the global symmetry, such as monopoles or the adjoint mass $\bar \Psi^i\Psi_j$, follows immediately from the broken symmetry according to the usual coset construction \cite{PhysRev.177.2247}.}
 \begin{equation}\bar \Psi^i \Psi_i\longleftrightarrow \epsilon^{abc}\text{Tr}(g^{-1}\partial_a gg^{-1}\partial_b gg^{-1}\partial_c g)~.\end{equation}

Eventually as we further increase the masses and reach $\theta=0\ {\rm mod}\ 2\pi$, we encounter $O(4)$ WF fixed points and hence an enhanced ``custodial'' symmetry. We can now identify $\bar \Psi_i \Psi^i$ with the unique relevant singlet of $O(4)$ WF \footnote{The monopole operators combine with non-monopole operators to form irreps of $O(4)$, and map to WF operators. For instance, the $q=2$ monopole combines with the adjoint mass $\bar \Psi^i\Psi_j$ to form the WF $O(4)$ rank 2 traceless symmetric operator, which is relevant.}, which has an accidental time-reversal symmetry in the IR that acts differently from that of massless QED$_3$. The summary of the phase diagram is in figure~\ref{FinalD}. While our discussion is somewhat tied to the concrete continuum theory~\eqref{kinetic}, the central ideas of how the $O(4)$ WF model could feature in the dynamics and how the anomalies and SPT phases could match (and the role of the $S^3$ nonlinear sigma model at $\theta=\pi$) should be more general and apply to various other systems in the same universality class (in particular certain lattice systems). 

For $N_f=3$, the time reversal breaking theory with $|k|=1/2$ may also have a symmetry breaking quantum phase similar to the $N_f=2$ scenario we discuss here. For $N_f\geq4$ and general $k$, however, it is expected to be just a standard second order phase transition \footnote{For $N_f=4$ and $k=0$, there is evidence of a second order phase transition both from lattice simulations \cite{2017PhRvX...7c1020H,2019PhRvL.123t7203H,2024PhRvX..14b1010W} as well as from bootstrap \cite{Albayrak:2021xtd,Chester:2016wrc}.}. 

It would be interesting to verify our scenario using a lattice simulation in the Villain formalism \cite{Sulejmanpasic:2019ytl,Gorantla:2021svj,Jacobson:2023cmr}, which allows one to suppress monopoles and thereby simulate the $SU(2)\times U(1)$ invariant theory. It would also be interesting to study the theory deformed by various monopole operators. 

{ While this paper was being finalized, we became aware of~\cite{Dumitrescu:2024jko}, where the same conclusion regarding the massless theory is reached via complementary arguments.}

\section{Weakly-Coupled Limits of QED$_3$} 

We denote the global symmetry gauge field which couples to the two fermions with opposite charge $\pm1$ by $X$ and the gauge field that couples to the monopole symmetry by $Y$. The dynamical gauge field is denoted by $a$ and it couples to the two Dirac fermions of charge 1.  
We take the Lagrangian to be 
\begin{equation}\label{masses} L_{kinetic}+
M_1 \bar \Psi_1\Psi_1+ M_2\bar \Psi_2 \Psi_2 +{1\over 2\pi} a dY+L_{ct.}~, \end{equation}
where $L_{kinetic}$ was given in~\eqref{kinetic} and $L_{ct.} = {1\over 4\pi } YdY-{1\over 4\pi} XdX$ represents the fact that we are adding some convenient counter-terms in the ultraviolet for the $X,Y$ background gauge fields.

Let us explain the reason for adding these particular counter-terms. The global structure of the gauge group that $X,Y$ couple to is in fact  
\begin{equation}\label{global} 
{U(1)\times U(1)\over \mathbb{Z}_2}~.
\end{equation}
The $\mathbb{Z}_2$ quotient is due to the fact that odd monopoles must be dressed by an odd number of fermions.
That means that in gapped phases the following counter-terms correspond to well-defined SPT phases
\begin{equation} 
\label{welldef}
{4k_1\over 4\pi}YdY~,\quad {4k_2\over 4\pi}XdX~,\quad {2k_3\over 4\pi}\left(YdY-XdX\right)~,
\end{equation}
with $k_{1,2,3}\in \mathbb{Z}$.   
The reason that we have picked the specific counter-terms in~\eqref{masses} is so that in the gapped phases below we obtain well-defined SPT phases.

When the masses $M_1,M_2$ are both large compared to $e^2$, the long distance limit can be easily understood by integrating out the fermions at one loop. 
\begin{itemize}
\item When both $M_1,M_2$ are negative and large we can integrate both fermions out and we obtain the low-energy theory
\begin{equation}-{1\over 4\pi}ada +{1\over 2\pi} a dY-{1\over 4\pi} XdX+L_{ct.}\longrightarrow {2\over 4\pi} YdY-{2\over 4\pi} XdX~,\end{equation}
where in the final step we have integrated out the dynamical gauge field $a$, which is at level 1 and hence the low-energy theory is invertible. We see that we have obtained a gapped phase with a well-defined SPT term, in accordance with~\eqref{welldef}.

\item When both $M_1,M_2$ are positive and large we obtain the low-energy theory
\begin{equation}{1\over 4\pi}ada +{1\over 2\pi} a dY+{1\over 4\pi} XdX+L_{ct.}\longrightarrow 0~,\end{equation}
where in the final step we have again integrated out the dynamical gauge field $a$. Again we have obtained a well-defined gapped phase.

\item When $M_1$ is large and positive while $M_2$ is large and negative we find the low-energy theory  
\begin{equation}\label{NGph}-{1\over 4e^2}F^2 +{1\over 2\pi} a d(X+Y)+{1\over 4\pi } YdY-{1\over 4\pi} XdX\,.\end{equation}
We have restored the kinetic term for the gauge field since there is no Chern-Simons term and thus the usual two-derivatives kinetic term dominates at long distances. This is a $U(1)\times U(1) \to U(1)$ symmetry breaking phase with a monopole condensate.  The spontaneously broken monopole symmetry couples to $X+Y$. Similarly, 
when $M_1$ is large and negative while $M_2$ is large and positive, we find  a symmetry breaking phase with a monopole condensate. Now the spontaneously broken symmetry couples to $X-Y$.

It looks alarming that the counter-term in~\eqref{NGph} is improperly quantized. However, this is a symmetry breaking phase with a Nambu-Goldstone boson due to monopole condensation. We will explain why we are allowed to add {\it certain} improperly quantized terms in this phase.

\end{itemize} 

The semi-classical limits can be seen in figure~\ref{FinalD}, where $M=M_1+M_2$, $A=M_1-M_2$. It is clear from the weakly-coupled limits that the physics at small $M_1,M_2$ must be nontrivial to accommodate these gapped SPT and gapless phases. It is also clear that it cannot be a single Landau-Ginzburg transition.  

{ The physics on the $\pm 45$ degrees lines that separate the four weakly-coupled quadrants can be readily understood. One Dirac fermion is massless on those lines. Since a $U(1)$ gauge field at level $\pm 1/2$ coupled to a single Dirac fermion is in the same universality class as the $O(2)$ Wilson-Fisher transition according to the dualities proposed in \cite{Barkeshli_2014,Chen_1993}, we conclude that each of the $\pm 45$ degrees lines represents a second-order phase transition in the $O(2)$ Wilson-Fisher universality class.}

\section{SPT Absorption}

We now address the Nambu-Goldstone phase~\eqref{NGph} in more detail, and in particular, the question about the normalization of the counter-term ${1\over 4\pi } YdY-{1\over 4\pi} XdX$ which is clearly incompatible with~\eqref{welldef}. 

We first prove that the partition functions of the $U(1)$ Nambu-Goldstone theory on manifolds with nontrivial $X+Y$ bundles vanish. For convenience denote $B=X+Y$ which is an ordinary, properly normalized $U(1)$ background gauge field and consider the theory
\begin{equation}\label{ANGB}-{1\over 4e^2}F^2 +{1\over 2\pi} a dB+\cdots~,\end{equation}
where $\cdots$ stand for possible higher derivative terms. The equations of motion in~\eqref{ANGB} dictate that $d  \star F\sim dB$, which by the Gauss law implies that $dB$ has to integrate to zero for the theory to have a nonzero partition function.

Going back to our case we see that our improperly quantized term ${1\over 4\pi } YdY-{1\over 4\pi} XdX={1\over 4\pi}(Y+X)d(Y-X) $ is linear in $Y+X$, which we argued is a trivial gauge field and hence the improper quantization raises no difficulties. 

The fact that the counter-terms do not have to be properly quantized in the NGB phase~\eqref{ANGB} does not mean that they are inessential. 
In the theory~\eqref{ANGB} 
there are defects which are electrically charged (we can represent them by Wilson lines). The counter-term ${1\over 4\pi } YdY-{1\over 4\pi} XdX$ leads 
to possibly fractional global charges of such electric defects. This is because in the presence of electric sources we can have nontrivial $B=X+Y$ bundles and the non-invariance of the partition function under gauge transformations of the background fields $X$ and $Y$  reflects the global symmetry charges of the electric sources.

We will later match these global charges as a consistency check on the phase diagram.

The whole discussion above can be phrased in the language of SPT absorption and symmetry fractionalization in a phase with one-form symmetry. This is carried out in the Appendix.

\section{The $O(4)$ Transition}
We now make some elementary comments about the $O(4)$ model deformed by quadratic terms that preserve $U(1)\times U(1)$. For convenience 
we assemble the four scalar fields $\phi_A=\phi_{1,2,3,4}$ into a  matrix as 
\begin{equation}\label{Zmatrix}\mathcal{Z} = \begin{pmatrix} Z_1 & -Z_2^* \\ Z_2 & Z_1^*\end{pmatrix}\end{equation} 
with $Z_1=\phi_1+i\phi_2$, $Z_2=\phi_3+i\phi_4$. 
We take the kinetic term to be as usual 
$|\partial Z_1+i YZ_1+i X Z_1|^2+|\partial Z_2+i YZ_2-i X Z_2|^2$, with $X,Y$ background gauge fields corresponding to a $U(1)\times U(1)$ symmetry. Note that the global structure is as before ${U(1)\times U(1)\over \mathbb{Z}_2}$.

\begin{figure}[!t]
  \includegraphics[width=0.49\textwidth ]{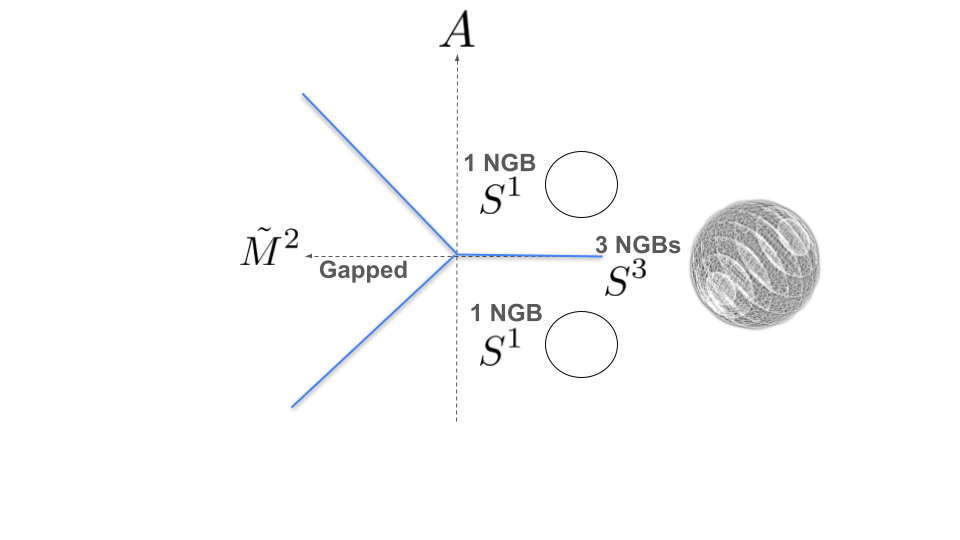}
  \caption{The phase diagram of the $O(4)$ Wilson-Fisher model deformed by mass terms preserving $U(1)\times U(1)$.}\label{phased}
\end{figure}

We take the potential to preserve $U(1)\times U(1)$:
$$V=\tilde M^2(|Z_1|^2+|Z_2|^2)-A(|Z_1|^2-|Z_2|^2)+\lambda (|Z_1|^2+|Z_2|^2)^2 ~.$$ 
The phase diagram can be readily obtained, see figure~\ref{phased}. { On the $\pm 45$ degrees lines in figure~\ref{phased} we have one massless complex scalar and hence the $O(2)$ Wilson-Fisher transition. This is exactly what we have found on the $\pm 45$ degrees lines in QED$_3$. }

Note that for $A>\tilde M^2$ and $A>0$ the broken symmetry couples to $X+Y$ since $Z_1$ condenses while for $A<-\tilde M^2$ and $A<0$ the broken symmetry couples to $X-Y$ since $Z_2$ condenses. In between, for $A=0$ there is $O(4)$ symmetry and the broken phase has 3 NGBs living on $S^3$. This sigma model with $S^3$ can be written as 
\begin{equation} \frac{f}{2} \int d^2xdt \ \delta^{ab}\text{Tr}(\partial_a g \partial_b g^{-1})~,\end{equation}
with $g$ an $SU(2)$ group element. Since it is the standard $O(4)$ transition, there is no $\theta$ angle.
A variant of the above discussion is to add the counter-term (properly quantized, compatible with~\eqref{welldef}) ${2\over 4\pi}YdY-{2\over 4\pi}XdX$ in the ultraviolet. This counter-term then survives in the infrared of the gapped phase for large positive $\tilde M^2$. 

Let us compare the broken phases we obtain here with those of QED$_3$,~\eqref{NGph}. 
In the $O(4)$ model we obtain the broken phase (when $Z_1$ condenses)
\begin{equation}\label{Abso} -{1\over 4e^2}F^2+{1\over2\pi} adY~. \end{equation}
On top of this, we may or may not add 
${2\over 4\pi}YdY-{2\over 4\pi}XdX$, depending if it was added in the ultraviolet. (Here we wrote the low-energy theory of the NGB in the $O(4)$ model using a dual $U(1)$ gauge field to make the comparison with QED$_3$ easier.) 

To make contact with QED$_3$ we need one last step, which is to match the fractionalization classes in the symmetry broken phases of the two models. 
In the $O(4)$ model in the phase with $Z_1$ condensed, the excitations on top of the vortex which carry charge under the unbroken symmetry are the $Z_2,Z_2^*$ particles, whose charge under the $X-Y$ gauge field is $\pm 1$. Therefore whether or not the we add the SPT term 
${2\over 4\pi}YdY-{2\over 4\pi}XdX$ the vortices carry integer charge under the unbroken symmetry.
Now consider the QED$_3$ model. In a background with unit background charge, we must excite, for instance a fermion, which carries $X-Y$ charge $\pm 1/2$, however, the counterterm in~\eqref{masses} makes the total unbroken charge into an integer and hence the quantum numbers of the defects agree.

\section{Enhanced Symmetry in QED$_3$}

QED$_3$ with $M_1=M_2$ in~\eqref{masses} preserves at short distances the continuous symmetry $SU(2)\times U(1)$. The phase diagram in figure~\ref{FinalD}, leads to symmetry enhancement to $O(4)$ at certain special values of the masses. We need to show that this enhancement is natural from the RG group point of view -- i.e. does not require additional tuning.

We consider Landau-Ginzburg models for the matrix~\eqref{Zmatrix}. We can act from the left and right by $SU(2)$ matrices
$\mathcal{Z}\to A \mathcal{Z} B^\dagger$, which induce $SO(4)={SU(2)_L\times SU(2)_R\over \mathbb{Z}_2}$ transformations of the components $\phi_{1,2,3,4}$. The determinant of $\mathcal{Z}$ is $\sum_A \phi_A^2$ which is why the $SU(2)_L\times SU(2)_R$ action induces $SO(4)$ transformations. We would like to prove that any $SU(2)\times U(1)$ preserving potential respects the full $SU(2)_L\times SU(2)_R$ symmetry. (This is related to the ``custodial'' symmetry in Higgs physics~\cite{Sikivie:1980hm}.) Indeed, to preserve $SU(2)_L$ we have to use the combination $\mathcal{Z}^\dagger \mathcal{Z}$. This a priori transforms in the adjoint+singlet of $SU(2)_R$, however, because of Bose symmetry only the singlet of $SU(2)_R$ is present. Therefore any $SU(2)\times U(1)$ invariant potential is fully $SU(2)_L\times SU(2)_R$ invariant. Therefore encountering the $O(4)$ Wilson-Fisher transition is entirely natural because all the $SU(2)\times U(1)$ invariant operators which break $SU(2)_L\times SU(2)_R$ are expected to be highly irrelevant. 
For instance, consider the matrix $ \mathcal{Z}^\dagger \partial_\mu\mathcal{Z} $. All its elements are $SU(2)_L$ invariant. For instance, consider the combination $|Z_1^*\partial_\mu Z_1+Z_2^*\partial_\mu Z_2|^2$. While it is irrelevant at the fixed point, it has a somewhat important effect in the symmetry broken phase. 
It modifies the metric on the $S^3$ target space.
Indeed, one can put a left-invariant metric on $S^3$ which is not right-invariant (it is only invariant under the Cartan on the right). This is a ``squashed'' $S^3$, which we anticipated by adding the squashing terms in~\eqref{CondM} 

\subsection{Time Reversal Symmetry}

Importantly, QED$_3$ at the massless point~\eqref{kinetic} is time reversal symmetric. The charge $\pm1$ monopoles are bosons in the doublet of $SU(2)$. Let us assemble the four monopole operators into a matrix, analogous to~\eqref{Zmatrix}:\begin{equation}\label{Mmatrix}\mathcal{M} = \begin{pmatrix} \mathcal{M}_1 & -\mathcal{M}_2^* \\ \mathcal{M}_2 & \mathcal{M}_1^*\end{pmatrix}\end{equation} 
The microscopic symmetries ${SU(2)_L\times U(1)_R\over \mathbb{Z}_2}$ of QED$_3$ act on this quartet of monopoles by $\mathcal{M}\to U \mathcal{M} V$
with $U\in SU(2)_L$ and $V=\left(\begin{matrix}
e^{i\phi} & 0 \\ 0 & e^{-i\phi}\end{matrix}\right)$.

There is also a microscopic charge conjugation symmetry which acts as $\mathcal{M}\to 
\mathcal{M}i \sigma_2$. This symmetry reverses the monopole number and extends the symmetry to ${SU(2)_L\times Pin^{-}(2)\over \mathbb{Z}_2}$, because charge conjugation is an element of order 4~\cite{Cordova:2017kue}.

Finally, the massless QED$_3$ point respects time reversal symmetry which acts on the quartet of monopoles as 
\begin{equation}\label{UVT}T_{UV}:\mathcal{M}(t)\longrightarrow i \sigma_2\mathcal{M}(-t)~.  \end{equation}
This definition satisfies $T_{UV}^2=-1$ on $\mathcal{M}$. More generally~\cite{Cordova:2017kue} $T_{UV}^2=(-1)^q$, with $q$ the monopole charge.
A virtue of this definition is that it commutes with $SU(2)_L$ because of the property of $SU(2)$ matrices 
$- i\sigma_2 U i\sigma_2 = U^*$ for $U\in SU(2)$. The anomalies for such time reversal symmetry were discussed in~\cite{Cordova:2023bja}.

In the ultraviolet theory, the mass deformation with $M_1=M_2$ is time-reversal odd. From the phase diagram in figure~\ref{FinalD} we see that the counter-term for the $U(1)\times U(1)\over \mathbb{Z}_2$ symmetry jumps, which indicates a mixed anomaly between $T_{UV}$ and ${U(1)\times U(1)\over \mathbb{Z}_2}$. It is therefore necessary for our proposal to match this anomaly at the massless point. 
Since in our scenario we find a condensate~\eqref{CondM}
and since~\eqref{UVT} satisfies $T_{UV}^2=-1$ on the elementary monopole, it follows that it is not possible to have a time-reversal invariant condensate of the elementary monopole. In other words, the time reversal symmetry is broken spontaneously in the $S^3$ nonlinear sigma model phase.

Even though the time reversal { symmetry} is spontaneously broken, we still need to match the anomaly (it is matched by the physics of various defects~\cite{Hason:2020yqf}). A related problem is that our discussion of the SPT absorption mechanism pertained to the {NGB} phase, which had a one-form symmetry. There is no one-form symmetry in the $S^3$ symmetry breaking phase since there are no vortices.

Therefore we have two remaining issues to address: How does the SPT phase jump as we move horizontally in the $S^3$ phase of figure~\ref{FinalD}? And how does { the} time reversal symmetry { anomaly} match at the middle point, $M=0$? Since the microscopic theory breaks time reversal symmetry at the points of the $O(4)$ { transition} by irrelevant operators, it is expected that in the broken phase a $\theta$ angle can be present. The $\theta$ angle is associated to $\pi_3(S^3)$, and we can add it to the sigma model
\begin{equation}{f\over 2} \int d^2xdt \ \delta^{ab}\text{Tr}(\partial_a g \partial_b g^{-1})+\theta \Gamma~,\end{equation}
with $\Gamma = {1\over 24\pi^2} \int d^2xdt\ \epsilon^{abc}Tr(g^{-1}\partial_a gg^{-1}\partial_b gg^{-1}\partial_c g)\in \mathbb{Z}$ is the winding number of space-time into the sigma model target space. An interesting fact is that as we change $\theta\to \theta+2\pi$, nothing happens in the bulk physics, but when we couple the system to the background fields $X,Y$, the partition function jumps precisely by ${2\over 4\pi }XdX-{2\over 4\pi }YdY$, as we show in the Appendix. Therefore we can account for the jump in the SPT phase by allowing $\theta$ to change continuously in the $S^3$ symmetry broken phase from $0 $ to $2\pi$, where we encounter our two $O(4)$ WF transitions. At the massless point of QED$_3$, we find precisely $\theta=\pi$ as required by time reversal symmetry. This reproduces the time reversal 't Hooft anomaly since $\theta=\pi$ is not quite time reversal invariant in the presence of background fields, analogously to the discussion in~\cite{Gaiotto:2017yup}. See also~\cite{Senthil:2005jk,Xu:2011sj} where it was emphasized that $\theta\to\theta+2\pi$ leads to edge modes, i.e. an SPT phase. In conclusion, the massless theory~\eqref{kinetic} flows to the $\pi=\theta$ sigma model with $S^3$ target space topology and as we deform by the mass term $M$ in the ultraviolet, $\theta$ changes and at $\theta=0,2\pi$ we encounter two WF $O(4)$ transitions, and as the mass is further increased, we obtain a gapped phase or a gapped SPT phase, depending on the sign of $M$.

\section*{Acknowledgments}

We thank 
Shailesh Chandrasekharan,  
Igor Klebanov, 
Max Metlitski, 
Silviu Pufu, 
Subir Sachdev, 
Sahand Seifnashri, 
Shu-Heng Shao, 
Nathan Seiberg 
Ning Su and 
Cenke Xu 
for useful conversations, and Nathan Seiberg, Sahand Seifnashri, and Cenke Xu for reviewing the manuscript.  SMC is supported by the UK Engineering and Physical Sciences Research council grant number EP/Z000106/1, and the Royal Society under the grant URF\textbackslash R1\textbackslash 221310. ZK is supported in part by the Simons Foundation grant 488657 (Simons Collaboration on the Non-Perturbative Bootstrap), the BSF grant no. 2018204, and the NSF award number 2310283.

\appendix

\section{Monopole operator scaling dimensions}

\begin{table}
\centering
\begin{tabular}{c||c|c|c|c|c}
$q$ & $ 2\Delta_{q}^{(0)} $ &  $\Delta_{q}^{(1)}$ &  $N_f=2$ & $O(4)$ & Error (\%)  \\ 
\hline
\hline
$1$&0.5302 &$ -0.038138$& 0.492062&{ 0.515(3)}&4.5\\
$2$&1.3463&$-0.19340(3)$&1.1529&{ 1.185(4)}&2.7\\
$3$&2.37286 &$-0.42109(4)$&1.95177&{ 1.989(5)}&1.9\\
$4$&3.5738 &$-0.70482(9)$&2.86898&{ 2.915(6)}&1.6 \\
$5$&4.9269 &$-1.0358(2)$&3.8911&{ 3.945(6)  }&1.4 \\
$6$&6.41674 &$-1.4082(2)$&5.00854&{ 5.069(7) }&1.2 \\
$7$&8.03182 &$-1.8181(2)$&6.21372&{ 6.284(8) }&1.1 \\
$8$&9.76308 &$-2.2623(3)$&7.50078&{ 7.575(9) }&1.0 \\
$9$&11.6032&$-2.7384(3)$&8.86482&{ 8.949(10) }&0.9 \\
$10$&13.5462 &$-3.2445(3)$&10.3017&{ 10.386(11) }&0.8 \\
\hline
\end{tabular}
\caption{Scaling dimensions $\Delta_{q}=N_f\Delta_{q}^{(0)}+\Delta_{q}^{(1)}+O(1/N_f)$ for charge $q$ scalar monopole operators in QED$_3$ in a large $N_f$ expansion \cite{Pufu:2013vpa,Dupuis:2021flq} extrapolated to $N_f=2$, compared to values of the rank $q$ traceless symmetric operators in the critical $O(4)$ model as computed from the lattice \cite{Banerjee:2019jpw}, along with the relative errors from the comparison. }
\label{tab}
\end{table} 

In the table \ref{tab} we compare monopole operator scaling dimensions as computed in a large $N_f$ expansion to subleading order \cite{Pufu:2013vpa,Dupuis:2021flq,Dyer:2013fja}, and operators in the $O(4)$ Wilson-Fisher theory as computed from lattice simulations \cite{Banerjee:2019jpw}.

\section{More on SPT Absorption}
\label{sec:detailsSPT}

We consider the theory
\begin{equation}\label{ANGBsm}-{1\over 4e^2}F^2 +{1\over 2\pi} a dB+\cdots~,\end{equation}
where $\cdots$ stand for possible higher derivative terms. 
The theory~\eqref{ANGBsm} has a $U(1)$ one-form symmetry due to the absence of charged particles (which leads to the conserved two-form current $\partial^\mu F_{\mu\nu}=0$), let us denote the corresponding $U(1)$ background two-form gauge field by $C$ and consider
\begin{equation}\label{ANGBTwo}-{1\over 4e^2}(F+C)^2 +{1\over 2\pi} a dB+\cdots~,\end{equation}
which is almost invariant under the gauge symmetry, zero-form symmetry, and one-form symmetry $a\to a+d\varphi+\lambda$, $B\to B+d\omega$, and $C\to C-d\lambda$, respectively. The non-invariance under $\lambda$ transformations represents the mixed anomaly between the zero-form and one-form symmetries, which is as usual rectified by adding a bulk term in 3+1 dimensions ${1\over 2\pi }\int_{\mathcal{M}_4} CdB$.
Now let us suppose the original NGB theory is coupled to a $B$ bundle which is nontrivial, i.e. there is a two-cycle with ${1\over 2\pi}\int dB = n \neq 0$ with some nonzero integer $n$. Then from the anomaly ${1\over 2\pi }\int_{\mathcal{M}_4} CdB$ it follows that the partition function { vanishes}, as argued on general grounds in~\cite{Delmastro:2022}.

The preceding proof that the partition function vanishes is a more formal version of the argument in the main text.

Using $B=X+Y$ and using the anomaly term, 
${1\over 2\pi }\int_{\mathcal{M}_4} CdB= {1\over 2\pi }\int_{\mathcal{M}_4} Cd(X+Y)$ we can remove the counter-term ${1\over 4\pi } YdY-{1\over 4\pi} XdX={1\over 4\pi}(Y+X)d(Y-X) $ altogether by setting $C={1\over 2}d(X-Y)$. This is essentially the notion of a fractionalization class~\cite{Delmastro:2022,n:2022tyl}. The NGB theory does not have excitations that break the one-form symmetry, but such exist in the full theory, and they can be assigned a (possibly fractional) charge under the unbroken $X-Y$ symmetry. The mechanism for SPT absorption that we have seen here is that the low-energy theory has a one-form symmetry which participates in a mixed anomaly. Analogous mechanisms operate in various other models where SPT absorption occurs, see for instance~\cite{verresen2017one,Ji:2019ugf,Dumitrescu:2023hbe}.

Stepping back for a second, our discussion of the fate of SPT phases in the $U(1)$ NGB theory can be viewed from a more general perspective.
It is a general fact that SPT phases can be absorbed in spontaneously broken phases. Consider a system in $d$ space-time dimensions with {\it discrete} 0-form symmetry $G$. Now assume that the symmetry $G$ is broken. If we turn on $G$ gauge fields on non-contractible cycles then the system has to react by creating domain walls. If we denote the tension of the domain wall by $T$ then the partition function is exponentially small 
\begin{equation} Z\sim e^{-T Vol_{d-1}}~,\end{equation} with  $Vol_{d-1}$ the volume of the transverse space to the gauge field one-form. In the large-volume limit one could say that the partition function goes to zero. Therefore on manifolds with $G$ background gauge fields, the partition function vanishes. In this situation, when the SPT phase is activated exactly on those manifolds where the partition function vanishes, we say that the SPT phase is absorbed.

\section{$\theta$ and SPT}
\label{sec:details}

Here we will show that the action \begin{equation}\label{WZ} \theta \Gamma~,\end{equation} with
$\Gamma = {1\over 24\pi^2} \int d^2xdt\ \epsilon^{abc}\text{Tr}(g^{-1}\partial_a gg^{-1}\partial_b gg^{-1}\partial_c g)\in \mathbb{Z}$ is not invariant under $\theta\to\theta+2\pi$ in the presence of  $U(1)\times U(1)$ background fields, rather, the path integral transforms by the SPT phase
${2\over 4\pi }XdX-{2\over 4\pi }YdY$. Our derivation follows closely the analysis in~\cite{Tanizaki:2018xto}. The strategy is to first make the action manifestly gauge invariant, via
\begin{equation}\label{replacement}g^{-1}\partial_a g \to g^{-1}D_ag =g^{-1}\left(\partial_ag+iX\sigma_3g+iYg\sigma_3\right)~. \end{equation}
There is no reason that the coupling to $X,Y$ should be minimal, and in fact since the replacement rule~\eqref{replacement} leads to linear terms in $X,Y$ upon expanding~\eqref{WZ}, it is convenient to consider deforming the action by the following non-minimal coupling
\begin{equation}\label{WZnonminimal} {i\over 4\pi}\epsilon^{abc}\text{Tr}\left(gD_ag^{-1}\partial_b X_c\sigma_3+g^{-1}D_ag\partial_b Y_c\sigma_3\right)~.\end{equation}
This non-minimal coupling is still manifestly gauge invariant.  
A short calculation shows that
\begin{align}\label{cal}&\epsilon^{abc}\int d^2xdt \text{Tr}\bigl[{1\over 12\pi} g^{-1}D_a gg^{-1}D_b gg^{-1}D_c g \nonumber\\ &
+  {i\over 4\pi}\left(gD_ag^{-1}\partial_b X_c\sigma_3+g^{-1}D_ag\partial_b Y_c\sigma_3\right)\bigr]\nonumber\\
  & =\int d^2xdt {1\over 12\pi} \epsilon^{abc}\text{Tr}\left(g^{-1}\partial_a gg^{-1}\partial_b gg^{-1}\partial_c g\right)
+ \nonumber \\ &\epsilon^{abc}{2\over 4\pi}\int d^2xdt \left(X_a\partial_bX_c-Y_a\partial_bY_c\right)~.   \end{align}
In summary, we find that upon $\theta\to\theta+2\pi$, the action transforms precisely by the required SPT phase.

\bibliographystyle{unsrt}
\bibliography{bibliography}
	
\end{document}